\documentclass[english,aps,prb,twocolumn,showpacs,showkeys]{revtex4}
\usepackage[T1]{fontenc}
\usepackage[latin1]{inputenc}
\usepackage{graphicx}

\makeatletter

\providecommand{\tabularnewline}{\\}
\newcommand{\lyxdot}{.}

\usepackage{epsfig}

\usepackage{prettyref}

\usepackage{babel}
\makeatother

\begin{document}

\title{Supercooling and the Metal-Insulator Phase Transition of NdNiO$_{3}$}

\author{Devendra Kumar}

\affiliation{Department of Physics, Indian Institute of Technology, Kanpur, 208016,
India}

\author{K. P. Rajeev}

\email{kpraj@iitk.ac.in}

\affiliation{Department of Physics, Indian Institute of Technology, Kanpur, 208016,
India}

\author{J. A. Alonso}

\affiliation{Instituto de Ciencia de Materiales de Madrid, CSIC, Cantoblanco,
E-28049 Madrid, Spain}

\author{M. J. Martínez-Lope}

\affiliation{Instituto de Ciencia de Materiales de Madrid, CSIC, Cantoblanco,
E-28049 Madrid, Spain}

\begin{abstract}
We report the temperature and time dependence of electrical resistivity
on high temperature, high oxygen pressure prepared polycrystalline
samples of NdNiO$_{3}$. NdNiO$_{3}$ is metallic above 195\,K and
below that temperature it undergoes a transition to an insulating
state. We find that on cooling NdNiO$_{3}$ below 195\,K it goes
into a state which is not in thermodynamic equilibrium and slowly
relaxes over several hours. As we cool it further and go below about
110\,K it goes into a stable insulating state. On heating the system
from the insulating state towards 200\,K we find that it remains
stable and insulating and undergoes a rather sharp insulator to metal
transition in the temperature range 185\,K to 195\,K. We try to
make sense of these and a few other interesting observations on the
basis of our current understanding of first order phase transitions,
supercooling, and metal-insulator transitions.
\end{abstract}



\maketitle

\section{INTRODUCTION}

Rare earth nickelates with the chemical formula RNiO$_{3}$, where
R stands for a rare earth ion, is one of the few families of oxides
that undergoes a first order metal-insulator phase transition. A metal-insulator
transition (M-I transition) is an electronic phase transition which
is usually solid to solid. The first order electronic phase transition
in many systems is associated with a hysteresis, and they exhibit
a phase separated (PS) state in the vicinity of the transition \citep{Granados_1,Chen,Chaddah}.
The PS state has drawn a lot of attention in the case of manganites,
mainly due to its implications for colossal magnetoresistance, and
it exhibits a variety of time dependent effects such as relaxation
of resistivity \citep{Chen,Levy} and magnetization \citep{Ghivelder}
and giant 1/f noise \citep{Podzorov}.

RNiO$_{3}$ crystallizes in the orthorhombically distorted perovskite
structure with space group P$_{bnm}$. The degree of distortion increases
with decrease in size of rare earth ion as one moves towards the right
in the periodic table \citep{Medarde}. The ground state of nickelates
(R $\neq$ La) is insulating and antiferromagnetic. On increasing
the temperature these compounds undergo a temperature driven antiferromagnetic
to paramagnetic transition, and an insulator to metal transition.
For NdNiO$_{3}$ and PrNiO$_{3}$, the M-I transition temperature
($T_{MI}$) and the magnetic ordering temperature ($T_{N}$) are the
same, while $T_{MI}$ is greater than $T_{N}$ for later members of
the series such as EuNiO$_{3}$ and SmNiO$_{3}$. The $T_{MI}$ of
nickelates increases with decreasing size of the rare earth ion and
the later members of the series also show a charge ordering transition
associated with the M-I transition \cite{Medarde,Alonso_1,Alonso_2}.
The charge ordered state has also been observed in epitaxially grown
thin films of NdNiO$_{3}$\citep{Staub}.

The M-I transition in NdNiO$_{3}$ is associated with a latent heat
and a sudden increase in unit cell volume which are signatures of
a first order phase transition\cite{Granados,Torrance}. There is
also a hysteresis in electronic transport properties just below $T_{MI}$
\citep{Granados}. The hysteresis in transport measurements points
toward the coexistence of metallic and insulating phases below $T_{MI}$
\cite{Granados}. In this paper we report a detailed study of the
nature of NdNiO$_{3}$ below $T_{MI}$ through temperature and time
dependent electrical resistivity measurements.

\section{EXPERIMENTAL DETAILS}

Polycrystalline NdNiO$_{3}$ samples in the form of 6\,mm diameter
and 1\,mm thick pellets were prepared and characterized as described
elsewhere\citep{Lcorre}. The preparation method uses a high temperature
of 1000$^{\circ}$C and a high oxygen pressure of 200\,bar.

All the temperature and time dependent measurements were done in a
home made cryostat. To avoid thermal gradients in the sample during
measurement it was mounted inside a thick-walled copper enclosure
so that during the measurement the sample temperature would be uniform.
It was found that mounting the sample like this improved the reproducibility
of the data, especially in the time dependence measurements, significantly.
A Lakeshore Cryotronics temperature controller model 340 was used
to control the temperature and the temperature stability was found
to be better than 3\,mK during constant temperature measurements. 

Below $T_{MI}$ NdNiO$_{3}$ is not in thermodynamic equilibrium and
slowly relaxes. Because of this the data we get depend on the procedure
used for the measurement. We used the following procedure to measure
the temperature dependence of resistivity. While cooling we start
from 300\,K, and then record the data in steps of 1\,K interval
after allowing the temperature to stabilize at each point. In between
two temperature points the sample was cooled at a fixed cooling rate
of 2\,K/min. After the cooling run is over we wait for one hour at
82\,K and then the heating data was collected at every one degree
interval. The intermediate heating rate between temperature points
was the same as the cooling rate used earlier. This cycle of measurements
was repeated with a different cooling and heating rate of 0.2\,K/min
also.

It was observed that the resistivity above $T_{MI}$ does not show
any time dependence, and it is also independent of measurement history.
Thus to avoid the effect of any previous measurements, all time dependent
experiments in the cooling run were done as follows: first take the
sample to 220\,K (above $T_{MI}$), wait for half an hour, then cool
at 2.0\,K/min to the temperature of interest and once the temperature
has stabilized record the resistance as a function of time. In the
heating run the time dependent resistivity was done in a similar fashion:
first take the sample to 220\,K, wait for half an hour, then cool
at 2.0\,K/min to 85\,K, wait for one hour, and then heat at 2.0\,K/min
to the temperature of interest and once the temperature has stabilized
record the resistance as a function of time.

The four probe van der Pauw method was used to measure the resistivity
and standard precautions, such as current reversal to take care of
stray emfs, were taken during the measurement. We also took care to
ensure that the measuring current was not heating up the sample. A
Keithley current source model 224 and a DMM model 196 were used for
the resistivity measurements.

\section{RESULTS}

Figure %
\begin{figure}[b]
\begin{centering}
\includegraphics[width=1\columnwidth]{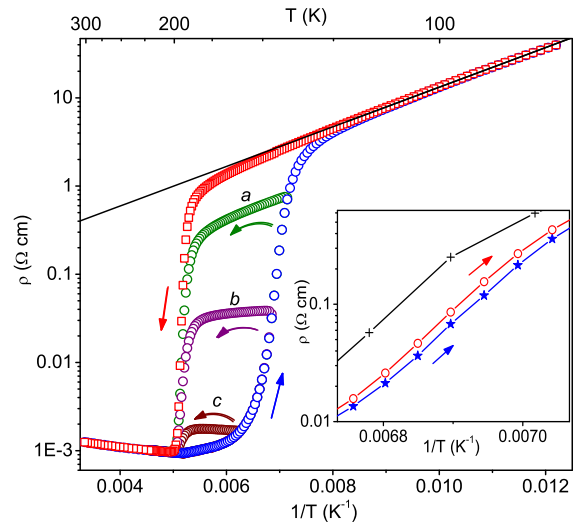} 
\par\end{centering}

\caption{(Color online) $\rho$ vs. $1/T$ plot for NdNiO$_{3}$. The blue
circles represent cooling data and the red squares stand for heating
data (cooling/heating rate 2.0 K/min). The solid line is a least square
fit to the band gap insulator model below 115\,K. The curves labeled
a, b \& c were taken as described in the text. The accuracy of the
data points is better than 1\% everywhere. The error does not exceed
0.8 m$\Omega$cm anywhere. The inset shows $\rho$ vs $1/T$ for three
different cooling rates: lower curve (blue stars): 2\,K/min, middle
curve (red circles): 0.2\,K/min, upper curve (black pluses): infinitely
slowly (explained in text). The connecting lines are to guide the
eyes.}

\label{fig:R vs T} 
\end{figure}
\ref{fig:R vs T} shows the electrical resistivity of NdNiO$_{3}$
as a function of temperature. We see that the function is multiple
valued, the cooling and heating data differing significantly from
each other and forming a large hysteresis loop. The resistivity plot
indicates that NdNiO$_{3}$ undergoes a relatively sharp M-I transition
at about 190\,K while heating with a width of about 10\,K. In contrast,
while cooling, the resistivity shows a rather broad M-I transition
centered around 140\,K with a spread of about 40\,K. Below 115\,K
or so, the heating and cooling data merge and the $\log\rho$ vs $1/T$
plot is linear. This indicates that the sample is insulating at low
temperatures and, if the band gap is $\Delta$, the resistivity should
follow the relation \begin{equation}
\rho(T)=\rho_{0}\exp(\Delta/k_{B}T)\label{eq:Band-gap-Resistivity}\end{equation}
Below 115\,K the data fit quite well to this model, with a coefficient
of determination, $R^{2}$= 0.99955. $\rho_{0}$ and $\Delta$ for
the insulating region turn out to be 99\,m$\Omega$\,cm and 42\,meV
respectively, which are in reasonable agreement with the values previously
reported\citep{Granados}.

We collected more hysteresis data with different minimum temperatures
such as 140\,K, 146\,K and 160\,K. In these measurements we cool
the sample from 220\,K to one of the minimum temperatures mentioned
above and then heat it back to 220\,K, both operations being carried
out at a fixed rate of 2\,K/min. Loops formed in this fashion are
called minor loops and these are indicated by the labels \emph{a},
\emph{b} and \emph{c} in Figure \ref{fig:R vs T}. In the cooling
cycle all the three minor loops coincide with the cooling curve of
the full hysteresis loop. In the heating cycle, for loops \emph{a}
and \emph{b }with lower minimum temperatures, the resistivity decreases
with increasing temperature and joins the full loop at 195\,K. In
the case of loop \emph{c,} as we increase the temperature, the resistivity
increases somewhat till about 182\,K and then it falls and joins
the full loop at 195\,K.

\begin{figure}[!t]
\begin{centering}
\includegraphics[width=1\columnwidth]{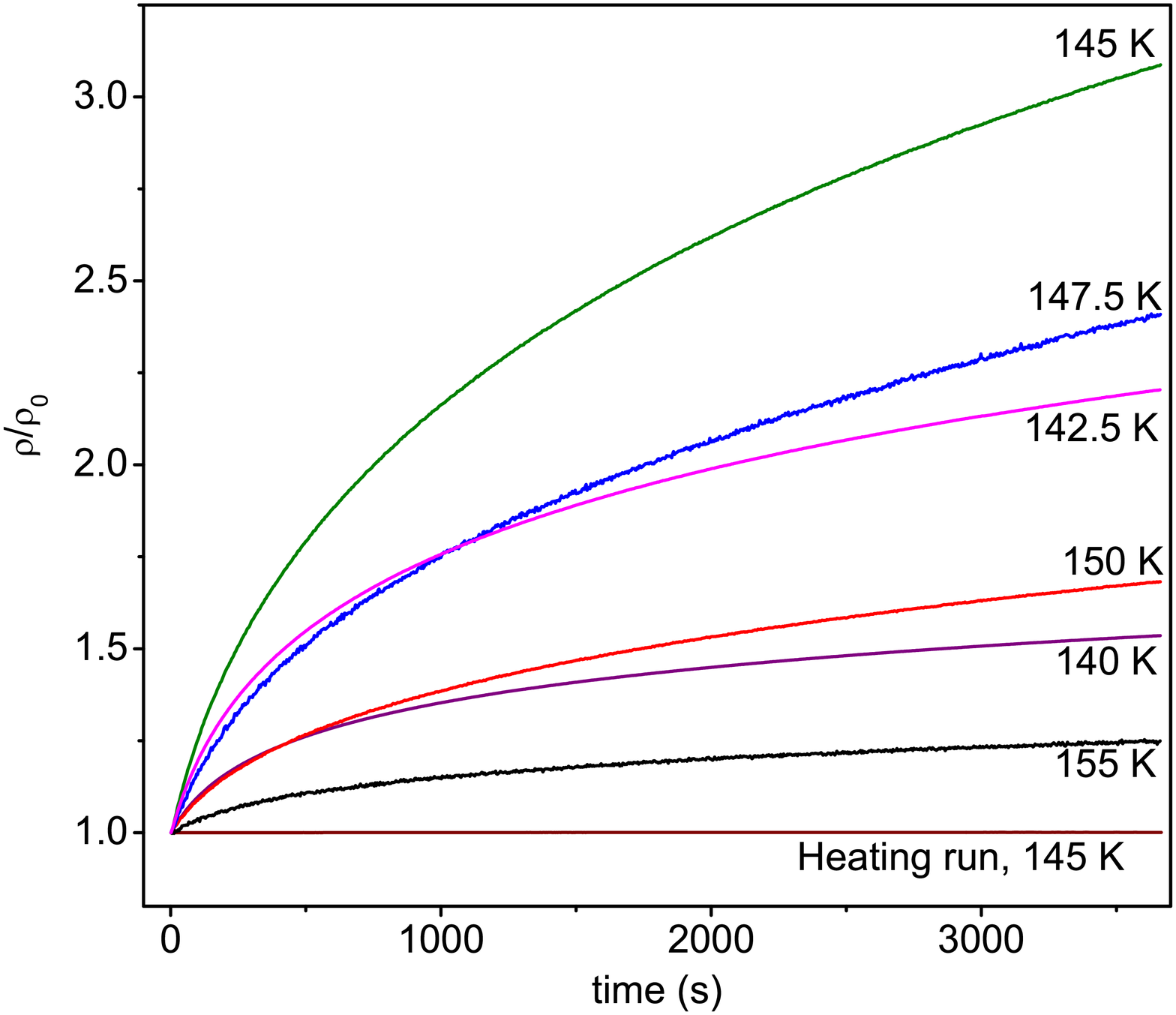} 
\par\end{centering}

\caption{(Color online) Time dependence of resistivity while cooling, at various
temperatures in the range 140 to 155\,K, for a period of one hour.
Maximum time dependence is seen at 145\,K which is about 300\,\%.
The curve at the bottom, which looks like a straight line, shows the
increase in resistivity in a heating run taken at 145\,K and the
change in this case is less than 0.2\,\%. \label{fig:time-dependance}
Not all the data are shown here to avoid clutter.}

\end{figure}

The resistivity also shows a noticeable dependence on the rate of
temperature change in the cooling cycle as shown in the inset of Figure
\ref{fig:R vs T}. The data for the lowest curve was collected at
2\,K/min and for the middle curve at 0.2\,K/min. The uppermost curve
is an estimate obtained by extrapolating the time dependence data
shown in Figure \ref{fig:time-dependance} to infinite time. We did
not see any rate dependence in the heating cycle. This is an indication
that while cooling, below the M-I transition temperature, the system
is not in equilibrium and hence the resistivity evolves with time.
The non-rate dependence observed in the data while heating tells us
that in the heating cycle the system is either in, or very close to,
equilibrium. These observations are corroborated by the data shown
in Figure \ref{fig:time-dependance}.

A subset of the time dependent resistivity data taken while cooling
is shown in Figure \ref{fig:time-dependance}. The data are presented
as $\rho(t)/\rho(t=0)$ so that the values are normalized to unity
at $t=0$ for easy comparison. We found that that below 160\,K, the
resistivity of the sample increases with time considerably. A maximum
relative increase in resistivity of about 300\,\% for a duration
of one hour is seen at 145\,K, the time dependence being lower both
above and below this temperature. We fitted the $\rho(T,t)$ curves
in figure 2 to the stretched exponential function \begin{equation}
\rho(t)=\rho_{0}+\rho_{1}(1-\exp(-(t/\tau)^{\gamma}))\label{eq:Stretched-Exponential}\end{equation}
where $\rho_{0}$, $\rho_{1}$, $\tau$ and $\gamma$ are fit parameters.
The fits are quite good with the $R^{2}$ value greater than 0.999
in most cases. See Table \ref{tab:Fit-parameters}. We note that the
exponent $\gamma$ lies in the range $0.5<\gamma<0.6$ and $\tau$
has a peak around 147.5\,K. The variation of resistivity with time
shows that the system slowly evolves towards an insulating state at
a constant temperature. We collected data up to 12 hours (not shown
here) to check whether the system reaches an equilibrium state, but
found that it was continuing to relax even after such a long time. 

\begin{table}
\begin{centering}
\begin{tabular}{|c|c|c|c|c|c|c|}
\hline 
\# & T(K) & $\rho_{1}/\rho_{0}$ & $\tau$ ($10^{3}$s) & $\gamma$ & $\chi^{2}/DOF$ & $R^{2}$\tabularnewline
\hline
\hline 
1 & 140.0 & 0.764(5) & 1.52(1) & 0.538(3) & 9.5 & 0.99971\tabularnewline
\hline 
2 & 142.5 & 1.89(1) & 2.02(2) & 0.554(2) & 5.9 & 0.99981\tabularnewline
\hline 
3 & 145.0 & 4.59(3) & 5.05(5) & 0.567(1) & 0.67 & 0.99993\tabularnewline
\hline 
4 & 147.5 & 3.39(6) & 7.9(3) & 0.568(3) & 0.51 & 0.99978\tabularnewline
\hline 
5 & 150.0 & 1.22(1) & 4.04(6) & 0.582(2) & 0.0014 & 0.99988\tabularnewline
\hline 
6 & 155.0 & 0.391(5) & 2.6(1) & 0.557(7) & 0.0003 & 0.99845\tabularnewline
\hline
\end{tabular}
\par\end{centering}

\caption{\label{tab:Fit-parameters}Fit parameters for the time dependence
data shown in Figure \ref{fig:time-dependance}. The degrees of freedom
of the fits $DOF\approx1000$. The $\chi^{2}/DOF$ for 150\,K and
155\,K are too small, indicating that we have overestimated the error
in resistivity in these cases. Anyway, we note that, the $R^{2}$
values are consistently good and indicate reasonably good fits.}

\end{table}

\begin{figure}[!t]
\begin{centering}
\includegraphics[width=1\columnwidth]{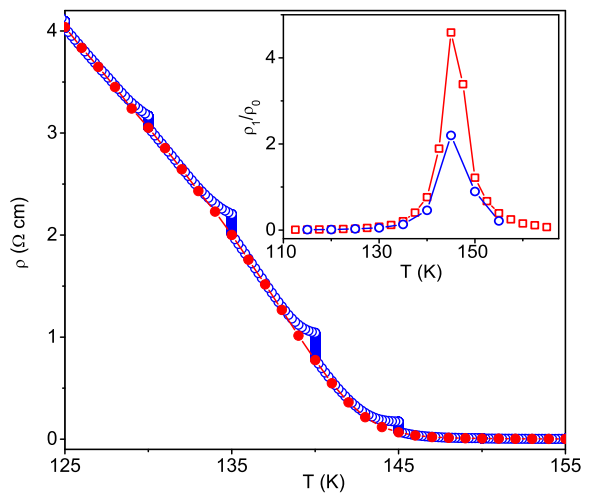} 
\par\end{centering}

\caption{(Color online) Temperature dependence of $\rho$ on cooling with ageing
by intermediate stops of one hour each (blue open circles) at 155,
150, 145, 140, 135, and 130 K. Red solid circles show the resistivity
without ageing. The inset compares the temperature dependence of $\rho_{1}/\rho_{0}$
calculated from the fit of Equation (\ref{eq:Stretched-Exponential})
to $\rho(T,t)$ data with intermediate ageing (blue circles) to the
time dependence data shown in Figure \ref{fig:time-dependance}(red
squares).}

\label{fig:res-bifurcation} 
\end{figure}

Figure \ref{fig:res-bifurcation} compares the resistivity in a cooling
run with and without intermediate aging. These data were taken as
follows: we start from 220\,K, come down to 160\,K at 2\,K/min,
collect time dependence data for one hour and after that resume cooling
at 2\,K/min and go down to 155\,K, collect resistivity time dependence
again for one hour and so on down to 110\,K for every 5\,K interval.
We note that when cooling is resumed after aging for an hour, $\rho(T)$
curve merges smoothly with the curve obtained without aging. A rather
similar observation has been reported in the phase-separated manganite
La$_{0.5}$Ca$_{0.5}$Mn$_{0.95}$Fe$_{0.05}$O$_{3}$ \citep{Levy}.

The red squares in the inset of Figure \ref{fig:res-bifurcation}
displays $\rho_{1}/\rho_{0}$ obtained from stretched exponential
fitting of time dependence resistivity data of Figure \ref{fig:time-dependance}.
The $\rho_{1}/\rho_{0}$ increases on decreasing the temperature below
160 K, attains a maximum around 145 K, and then decreases on further
lowering the temperature. We have already seen this behavior in Table
\ref{tab:Fit-parameters}. The value of $\rho_{1}/\rho_{0}$ obtained
from the fitting of time dependence data with intermediate aging is
shown as blue circles in the inset of figure \ref{fig:res-bifurcation},
and we note that the value of $\rho_{1}/\rho_{0}$ is relatively small
compared to what one obtains without intermediate aging.

In the heating cycle, the $\rho$ vs $T$ exhibits a rather sharp
insulator to metal transition. The magnitude of time dependence in
heating runs (maximum $\approx0.2\%$) is negligible compared to what
one gets in cooling runs (maximum$\approx300\%)$ (figure \ref{fig:time-dependance}),
which suggest that in the heating run, below $T_{MI}$, the sample
is almost fully insulating and stable.

The major observations that have been made so far are summarized below.

\begin{enumerate}
\item The metal to insulator transition in cooling runs is rather broad
and extends from about 160\,K to 120\,K.
\item In cooling runs a large amount of time dependence in resistivity is
seen compared to which the time dependence seen in heating runs is
negligible.
\item NdNiO$_{3}$ undergoes a rather sharp insulator to metal transition
between 185K and 195\,K while heating. 
\item There is a large hysteresis between the cooling and heating data below
the MI transition temperature.
\end{enumerate}

\section{DISCUSSION}

It has been known that the M-I transition in NdNiO$_{3}$ is associated
with a latent heat and a jump in the unit cell volume which are the
characteristics of a first order phase transition\cite{Torrance,Granados}.
DSC measurements of Granados et al. show that, while cooling, the
latent heat released during the M-I transition extends over a broad
temperature range\citep{Granados}. This is consistent with the results
of the resistivity measurements that the M-I transition in NdNiO$_{3}$
is broadened while cooling . Thus we can say that NdNiO$_{3}$ has
a broadened first order metal to insulator phase transition while
cooling.

We have seen that, below $T_{MI}$, while cooling, the system is not
in equilibrium and evolves with time which suggests that it is in
a metastable state. The resistivity of this metastable state slowly
increases with time which means that a slow metal to insulator transition
is going on in the system. On the other hand while heating up from
low temperature we saw that the system remains insulating all the
way up to 185\,K and then it transitions to the metallic state by
195\,K. The resistivity was found to have negligible time dependence
during the whole of the heating run. This result indicates that the
low temperature insulating state is stable and is probably the ground
state of the system. It is well known that below a phase transition
temperature a high temperature phase can survive as a metastable supercooled
state. Based on these facts we propose that the metastable state in
the case of NdNiO$_{3}$ consists of supercooled (SC) metallic phases
and stable insulating phases. 

In a first order transition, a metastable SC phase can survive below
the first order transition temperature ($T_{C}$), till a certain
temperature called the limit of metastability ($T^{*}$) is reached
\cite{Chaikin,Chaddah,Chaddah1}. In the temperature range $T^{*}<T<T_{C}$
there is an energy barrier separating the SC phase from the stable
ground state. The height of the energy barrier ($U$) separating the
SC metastable phase from the stable ground state can be written as
$U\propto f(T-T^{*})$, where $f$ is a continuous function of ($T-T^{*}$),
and vanishes for $T\leq T^{*}$. As the temperature is lowered, at
$T=T^{*}$, the SC metastable phase becomes unstable and switches
over to the stable ground state \citep{Chaikin}. At $T>T^{*}$ the
SC metastable phase can cross over to the stable ground state with
a probability (\emph{p}) which is governed by the Arrhenius equation
\begin{equation}
p\propto exp(-U/\mbox{k}_{B}T)\label{eq:Arrhenius}\end{equation}
which tells us that the barrier will be crossed with an ensemble average
time constant $\tau\propto1/p$. If we imagine an ensemble of such
SC phases with the same barrier $U$, then the volume of the metastable
phase will exponentially decay with a time constant $\tau$.

The energy barrier $U$ that we talked about in the previous paragraph
is an extensive quantity. Our system is polycrystalline, ie., it is
made up of tiny crystallites of different sizes. Each one of the crystallites
will have an energy barrier proportional to its volume. Thus for a
crystallite the energy barrier separating the metastable state from
the ground state can be written as \begin{equation}
U\propto Vf(T-T^{*})\label{eq:Barrier-height}\end{equation}
where $V$ is the volume of the crystallite. As already mentioned
$f$ is a continuous function which vanishes for non-positive values
of its argument. This means that various crystallites with the same
$T^{*}$ will have different energy barriers depending on their size
which implies that the time constant $\tau$, of the previous paragraph,
will spread out and become a distribution of time constants depending
on the distribution of the size of the crystallites. This can give
rise to the volume of the metallic state decaying in a stretched exponential
manner with time\cite{Palmer,Ediger}. This behavior of the metallic
volume with time will give rise to the resistivity also evolving with
time in a similar fashion. We shall see later in the discussion how
the metallic volume, the resistivity, and their time dependences are
connected with each other.

Imry and Wortis have argued that a weak disorder can cause a distribution
of $T^{*}$'s in different regions of a sample\cite{Imry,Chaddah}.
This has been experimentally observed in a vortex lattice melting
experiment in a high $T_{C}$ superconductor \cite{Soibel}. In our
case grain boundaries, shape of the crystallite etc. could be sources
of disorder which would give rise to different $T^{*}$'s for different
crystallites. This effect will further broaden the distribution of
time constants in our system which arise from the distribution of
crystallite sizes. 

Coming back to the heating runs we note that the lack of time dependence
of the resistivity data indicates that the system is in a stable state.
Now the question arises: what is the reason for the width of about
10\,K observed in the insulator to metal transition? Could it be
a superheating effect? It cannot be because, as already noted, the
lack of time dependence during heating runs rules out the possibility
of any metastable phase in the system. The broadening while heating
is most likely caused by the rounding of the phase transition due
to finite size of the crystallites. A similar observation on broadening
of phase transitions due to finite size of grains have been reported
in a CMR manganite\cite{Fu}.

Now we understand the reason for the hysteresis seen in this system.
During a heating run from low temperature the system is insulating
and hence its resistivity is high. While cooling some of the crystallites
transform themselves into the stable insulating state while others
remain in the supercooled metallic state . The presence of the metallic
crystallites lowers the resistivity of the system. Thus we get different
values of resistivity during heating and cooling runs giving rise
to hysteresis in resistivity.

The picture so far: NdNiO$_{3}$ is polycrystalline and thus consists
of tiny crystallites. Below $T_{MI}$, while cooling, the tiny crystallites
can be in either a supercooled metallic state or a stable insulating
state. A crystallite in the supercooled metallic state can make a
transition to the stable insulating state if it crosses the barrier
$U$ with the help of energy fluctuations which in our case are thermal
in origin. Each crystallite would undergo the transition independent
of each other. This is supported by the fact that the resistivity,
and as we shall see later, the volume of the metastable metallic state,
behave in a stretched exponential manner with time. At a particular
temperature, as a function of time, the transition from the metallic
to the insulating phase will begin with the smaller crystallites,
because the barrier is small for them, and then proceed with the larger
crystallites, in the increasing order of barrier size. While heating
the system remains insulating and stable all the way up to the M-I
transition and then it goes over to the stable metallic state. There
is no superheating during heating runs.

Now that we have understood the origin of time dependence and hysteresis
in the system we would like to consider some more issues that needs
to be understood. These are

\begin{enumerate}
\item The dependence of resistivity on cooling rate as shown in the inset
of Figure \ref{fig:R vs T}. 
\item The behavior of the minor loops seen in Figure \ref{fig:R vs T}.
\item The intermediate ageing behavior seen in Figure \ref{fig:res-bifurcation}.
\item Both $\rho_{1}/\rho_{0}$ and $\tau$ of Equation (\ref{eq:Stretched-Exponential})
go through a peak around 145\,K and 147.5\,K respectively as shown
in Table \ref{tab:Fit-parameters} and also in the inset of Figure
\ref{fig:res-bifurcation} and Figure \ref{fig:tau-vs-T}. 
\end{enumerate}
Let us take up these issues one by one.

\subsection{Cooling Rate Dependence}

In the inset of Figure \ref{fig:R vs T} we saw that while cooling
a lower cooling rate increases the resistivity of the sample in the
hysteresis region. A lower cooling rate allows a larger number of
the metastable metallic crystallites to switch over to the insulating
state and thus the resistivity will be larger if the measurement is
done slowly.

If the measurement is done infinitely slowly in the cooling run we
should get the uppermost curve shown in the inset of Figure \ref{fig:R vs T}.
It is interesting to note that this curve does not come anywhere near
the heating run curve. To paraphrase, even if we sit for an infinite
time on the cooling curve we cannot reach the heating curve. The reason
for this is that only a small number supercooled crystallites which
have a relatively small energy barrier are able to switch over to
the insulating state at any given temperature. The others with the
larger energy barriers remain trapped in the metallic state. These
crystallites with the larger barriers can switch their state only
if the temperature is lowered such that $T$ comes sufficiently close
to $T^{*}$ and thus the barrier becomes small and easy to cross (Equation
(\ref{eq:Barrier-height})).

\subsection{Minor Loops}

The minor loops \emph{a, b},\emph{ }and\emph{ c }shown in Figure \ref{fig:R vs T}
follow the full hysteresis loop in the cooling cycle till we start
the heating. While heating, curve \emph{a,} with the lowest minimum
temperature (140\,K), shows a falling resistivity with increasing
temperature. Curve \emph{b} with the intermediate minimum temperature
(146\,K) shows a rather flat resistivity with increasing temperature,
while curve \emph{c }with the highest minimum temperature (160\,K)
shows a rising resistivity with increasing temperature initially before
falling and joining the full hysteresis loop at the MI transition
temperature 195\,K.

The behavior of the minor loops can be understood by recognizing that
the nature of resistivity, when heating is resumed, in the metastable
region is determined by the competition between the temperature dependence
and time dependence of resistivity. In the case of curve \emph{a,
}with the lowest minimum temperature, most of the crystallites will
be in the insulating state as can be inferred from the large value
of resistivity. Only few of the crystallites will be in the metastable
state and hence the time dependence effects can be expected to be
small. Thus during heating the resistivity of the sample will be dominated
by the temperature dependence of resistivity of the insulating crystallites.
This is the reason the heating part of the curve looks more or less
parallel to the main heating curve before the onset of the transition.

The behavior of curve \emph{c }is easier to understand and hence let
us consider it before discussing curve \emph{b. }In this case the
resistivity is low and most of the crystallites are in the metallic
state. The metallic crystallites form a percolating network and the
resistivity of the sample will be governed by their behavior. The
insulating crystallites are essentially shorted out and will have
little role in determining the resistivity. The metallic crystallites,
being metastable, will switch to the insulating state governed by
Equation (\ref{eq:Arrhenius}), thus slowly increasing the resistivity
of the sample as a function of time. When the resistivity values are
collected while heating, there will be some time difference between
two data points depending on the heating rate, allowing a rise in
resistivity to register. We would then find that the resistivity increases
with increasing temperature. Curve \emph{b }lies between curves \emph{a
}and \emph{c }and in this case there will be a good number of both
insulating and metallic crystallites. Here there would be competition
between the temperature dependence of the insulating crystallites
and the time dependence of the metallic crystallites and hence one
can expect that the curve \emph{b} would be more or less flat at the
onset of heating.

\subsection{Intermediate Ageing}

Now let us consider the intermediate ageing data shown in Figure \ref{fig:res-bifurcation}.
We note that after ageing for one hour, when the cooling is resumed,
the slope of the the $\rho$ vs. $T$ curve is less in magnitude than
for the curve obtained without ageing. When we stop the cooling and
age the sample at a fixed temperature the supercooled metallic crystallites
with relatively small $U$ will switch over to the insulating state.
This decreases the number of crystallites in the metallic state and
raises the resistivity. So now, when the cooling is resumed, the number
of metallic crystallites which can switch to the insulating state
would be less and hence the change in resistivity with temperature
would be lower.

From Figure \ref{fig:res-bifurcation} it is clear that on resuming
cooling after an intermediate stop of one hour the cooling curve merges
with the curve obtained without ageing within about 3\,K or less.
As already noted in the previous paragraph when we stop the cooling
and age the sample at a fixed temperature the supercooled metallic
crystallites with relatively small $U$ will switch over to the insulating
state. As can be inferred from Equation (\ref{eq:Barrier-height})
those crystallites will have a small $U$ which have (i) their metastability
temperature ($T^{*}$) close to their temperature ($T$) (ii) a small
size. The major contributions to resistivity change will come from
the relatively larger crystallites making the transition from the
metallic state to the insulating state. In the light of this the merger
of the cooling curves with and without ageing within a small temperature
change of about 3\,K means that most of the crystallites which undergo
the transition from metal to insulator during ageing have their $T^{*}$
within a few kelvin of the temperature of the sample. We had already
reached this conclusion earlier when we discussed the cooling rate
dependence.

\subsection{Peaks in $\rho_{1}/\rho_{0}$ and $\tau$}

We had noted earlier that both $\rho_{1}/\rho_{0}$ and $\tau$ go
through peaks around 145\,K and 147.5\,K respectively. The peak
in $\rho_{1}/\rho_{0}$ is not surprising because it shows up in a
region of temperature in the cooling curve where the resistivity changes
very fast with temperature. Thus one would expect that a large number
of metastable crystallites are switching over to the insulating state
near this region. Hence if a relaxation measurement is done here we
should see the effect of a large number of crystallites transitioning
to the insulating state as a peak in $\rho_{1}/\rho_{0}$. But the
peak in $\tau$ at about 147.5\,K is not that easily understood.
To check whether this peak has anything to with percolation we should
examine how the metallic and insulating volumes depend on temperature.

The conductivity of a phase separated system depends on the volume
fraction of insulating and metallic phases, their geometries, distribution,
and their respective conductivities ($\sigma_{I}$ and $\sigma_{M}$).
McLachlan has proposed an equation based on a general effective medium
(GEM) theory for calculating the effective electrical conductivity
$\sigma_{E}$ of a binary MI mixture \citep{McLachlan}, which is 

\begin{equation}
(1-f)\frac{(\sigma_{I}^{1/t}-\sigma_{E}^{1/t})}{(\sigma_{I}^{1/t}+A\sigma_{E}^{1/t})}+f\frac{(\sigma_{M}^{1/t}-\sigma_{E}^{1/t})}{(\sigma_{M}^{1}/t+A\sigma_{E}^{1/t})}=0\label{eq:GEM}\end{equation}
where $f$ is the volume fraction of the metallic phases and $A=(1-f_{c})/f_{c}$,
$f_{c}$ being the volume fraction of metallic phases at the percolation
threshold, and $t$ is a critical exponent which is close to 2 in
three dimensions \citep{Herrmann,Hurvits}. The constant $f_{c}$
depends on the lattice dimensionality, and for 3D its value is 0.16
\citep{Efros}. The GEM equation has been successfully applied to
a wide variety of isotropic, binary, macroscopic mixtures and it works
well even in the percolation regime \citep{McLachlan,Hurvits,Kim}.

\begin{figure}[!t]
\begin{centering}
\includegraphics[width=1\columnwidth]{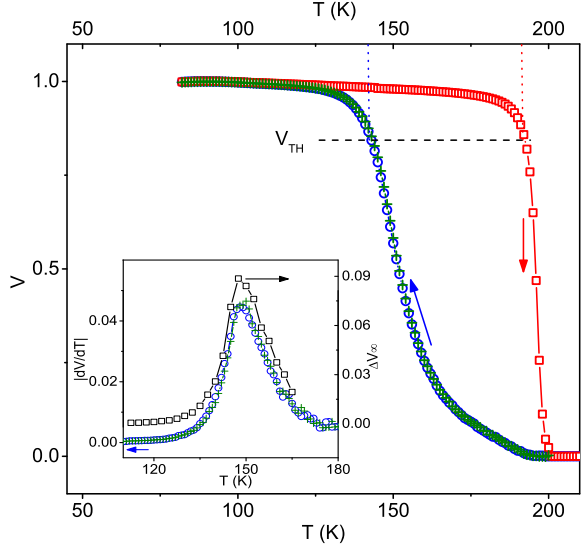} 
\par\end{centering}

\caption{(Color online)Temperature variation of $V$ ($=1-f$), the insulating
volume fraction. The blue circles show the cooling cycle (2\,K/min
cooling rate) and the red squares represent the heating cycle. The
green pluses represent the insulating volume fraction for 0.2\,K/min
cooling rate. The dashed horizontal line at 84\% insulating volume
represents the percolation threshold. The inset shows the variation
of $|dV/dT|$ (blue circles: cooling rate 2\,K/min, and green pluses:
cooling rate 0.2\,K/min) and $\Delta V_{\infty}$ (black squares)
with temperature. }

\label{fig:v-fraction} 
\end{figure}

In order to calculate the volume fraction of metallic and insulating
phases from Equation (\ref{eq:GEM}), we need their respective resistivities
$\rho_{M}(T)$ and $\rho_{I}(T)$. $\rho_{M}(T)$ was obtained using
$\rho_{M}=\rho_{0}+AT$, where $A$ is the temperature coefficient
of resistivity estimated from the resistivity data above the M-I transition.
$\rho_{I}(T)$ was calculated using the parameters obtained by fitting
the resistivity data below 115\,K to Equation (\ref{eq:Band-gap-Resistivity}).
Using the above information and the resistivity data, we calculated
the volume fraction of metallic and insulating phases and it is shown
in Figure \ref{fig:v-fraction}. In the cooling cycle the volume fraction
of the insulating phases ($V=1-f$) slowly increases on decreasing
the temperature below $T_{MI}$, while in the case of heating cycle
it remains nearly constant up to about 185\,K, and then drops to
zero by 200\,K. As can be seen from Figure \ref{fig:v-fraction}
the percolation threshold for the cooling runs occurs at around 144\,K.
Below this temperature there will be no continuous metallic paths
in the system.

\begin{figure}[!t]
\begin{centering}
\includegraphics[width=1\columnwidth]{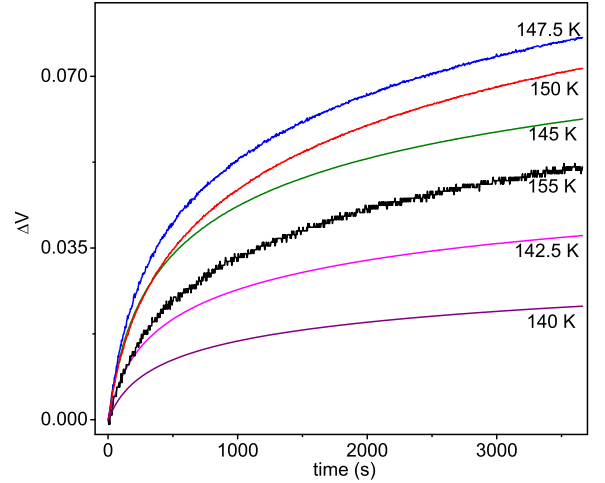} 
\par\end{centering}

\caption{(Color online) Increase in volume fraction of insulating phases ($\Delta V$)
with time, at various temperatures. }

\label{fig:v-time} 
\end{figure}

\begin{figure}[!t]
\begin{centering}
\includegraphics[width=1\columnwidth]{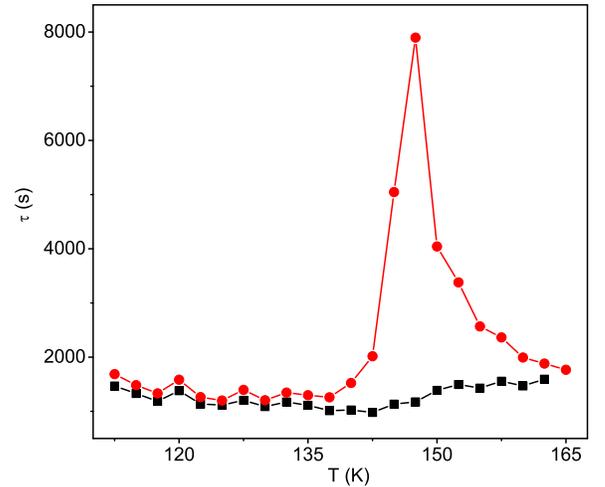} 
\par\end{centering}

\caption{(Color online)Temperature variation of $\tau$ obtained from the stretched
exponential fitting of $\rho(t)$ (red circles) and $\Delta V(t)$
(black squares) curves. }

\label{fig:tau-vs-T} 
\end{figure}

The rate of change of metallic volume fraction ($-dV/dT$) has a maximum
around 147.5 K in the cooling runs (inset of Figure \ref{fig:v-fraction}).
Figure \ref{fig:v-time} displays the increment ($\Delta V$) of insulating
volume fraction with time, that has been extracted from the time dependent
resistivity data of Figure \ref{fig:time-dependance}. Just as in
the case of time dependence of resistivity, the $\Delta V(t)$ curves
of Figure \ref{fig:v-time} were fitted to a stretched exponential
function $\Delta V(t)=\Delta V_{\infty}(1-\exp(-(t/\tau)^{\gamma}))$,
where the constant $\Delta V_{\infty}$ gives the increase in the
insulating volume fraction when waiting for an infinitely long time.
The inset of Figure \ref{fig:v-fraction} also shows the variation
of $\Delta V_{\infty}$ with temperature, which mimics the behavior
of $|dV/dT|$. Figure \ref{fig:tau-vs-T} compares the constant of
relaxation $\tau$ obtained from the fitting of $\rho(t)$ and $\Delta V(t)$
curves of Figure \ref{fig:time-dependance} and \ref{fig:v-time}
respectively. For the $\rho(t)$ curves, $\tau$ has a peak around
147.5 K while for the $\Delta V(t)$ curves, surprisingly, $\tau$
is nearly constant.

During cooling $|dV/dT|$ of the inset of Figure \ref{fig:v-fraction}
represents the amount of volume that will change from the supercooled
metallic to the insulating state for a unit temperature change. Now
a small change in temperature will result in those supercooled crystallites
with their $T^{*}$ falling in that temperature range switching to
the insulating state. This means that the value of $|dV/dT|$ at $T$
is a good measure of the fraction of supercooled metastable crystallites
which have their $T^{*}$ close to $T$. Thus $|dV/dT|$ of the cooling
curve represents the distribution of $T^{*}$'s in the system. In
coming to the above conclusion we have disregarded the small fraction
of crystallites that would be switching due to the time elapsed in
covering the small temperature change. This is supported by the fact
that the $|dV/dT|$ values calculated from the 2\,K/min and 0.2\,K/min
cooling curves are essentially the same.

The peak in $\triangle V_{\infty}$ occurs at 147.5\,K which is also
where the distribution of $T^{*}$'s peak. This is understandable
because all the crystallites with $T^{*}$ close to 147.5\,K would
be switching to the insulating state if we wait for a sufficiently
long time. Since the maximum number of crystallites with $T^{*}$
are close to 147.5\,K, we will get the maximum temperature dependence
for $\triangle V_{\infty}$ at 147.5\,K.

One would have expected that the peak in $\rho_{1}/\rho_{0}$ and
$\triangle V_{\infty}$ would occur at the same temperature. The peak
in $\rho_{1}/\rho_{0}$ is at 145\,K which is 2.5\,K below the peak
of $\triangle V_{\infty}$. The reason for this is that the resistivity
would change faster for a given metallic volume change nearer the
percolation threshold. We guess that even if the volume change is
smaller at 145\,K a larger resistivity change due to the proximity
to the percolation threshold compensates for the smaller volume change
and shifts the maximum of $\rho_{1}/\rho_{0}$ to that temperature.

As can be seen from Figure \ref{fig:tau-vs-T} the parameter $\tau$
of the resistivity time dependence fit has a peak around 147.5\,K
while there is no such peak in the $\tau$ of the volume fraction
time dependence. The rather flat temperature dependence of the $\tau$
of the volume fraction is probably an indication that the distribution
of energy barriers look more or less the same at all temperatures.
We note that the difference in behavior of the $\tau$'s from resistivity
and volume fraction is most pronounced close to the percolation threshold,
while away from the threshold both $\tau$'s behave in a very similar
manner. In a resistivity relaxation measurement, close to the percolation
threshold, the resistivity will change significantly even when the
metallic volume changes very little. Thus, one will find that, near
the percolation threshold, the resistivity will relax over a longer
time period.

\section{Conclusion}

Our experimental results demonstrate that while cooling the physical
state of NdNiO$_{3}$ is phase separated below the M-I transition
temperature; the phase separated state consists of SC metallic and
insulating crystallites. A metastable metallic crystallite switches
from the metallic to the insulating state probabilistically depending
on the closeness of the temperature of metastability and the size
of the crystallite. At low temperature, below 115\,K or so, the system
is insulating, all the crystallites having switched over to the insulating
state. While heating the crystallites remain in the stable insulating
state till the MI transition temperature is reached and then they
switch over to the metallic state.

\begin{acknowledgments}
DK thanks the University Grants Commission of India for financial
support during this work. JAA and MJM-L acknowledge the Spanish Ministry
of Education for funding the Project MAT2007-60536 
\end{acknowledgments}

\end{document}